\documentclass[aps,twocolumn,prl,showpacs,superscriptaddress,groupedaddress]{revtex4}

\usepackage{dcolumn}% Align table columns on decimal point
\usepackage{bm}% bold math
\usepackage{color}

\def\ltape{\hbox{\ $<$\hskip -8pt\raise -4pt\hbox{$\sim$}\ }}
\def\gtape{\hbox{\ $>$\hskip -8pt\raise -4pt\hbox{$\sim$}\ }}

\pdfoutput=1

   \ifx\pdfoutput\undefined
   \usepackage[dvips]{graphicx}
   \else
   \usepackage[pdftex]{graphicx}
   \fi
   \usepackage{epstopdf}
   \DeclareGraphicsExtensions{.pdf,.gif,.jpg,.eps,.png}
  
   \usepackage{amsmath}
  \usepackage{amssymb}
\begin{document}

%\preprint{{\it Physical Review Letters}}

\title{Suppression of collisionless magnetic reconnection in asymmetric current sheets}
%\title{Drift and suppression of collisionless magnetic reconnection in asymmetric current sheets}

\author{Yi-Hsin~Liu}
\affiliation{NASA-Goddard Space Flight Center, Greenbelt, MD 20771}
\author{Michael~Hesse}
\affiliation{NASA-Goddard Space Flight Center, Greenbelt, MD 20771}
\date{\today}

\date{\today}

\begin{abstract}

Using fully kinetic simulations, we study the suppression of asymmetric reconnection in the limit where the diamagnetic drift speed $\gg$ Alfv\'en speed and the magnetic shear angle is moderate. We demonstrate that the slippage between electrons and the magnetic flux facilitates reconnection, and can even result in fast reconnection that lacks one of the outflow jets. Through comparing a case where the diamagnetic drift is supported by the temperature gradient with a companion case that has a density gradient instead, we identify a robust suppression mechanism. The drift of the x-line is slowed down locally by the asymmetric nature of the current sheet and the resulting tearing modes, then the x-line is run over and swallowed by the faster-moving following flux. 

\end{abstract}


\maketitle

%{\color{red} \it Introduction--}
{\it Introduction--}
The explosive release of magnetic energy through magnetic reconnection is ubiquitous in laboratory, space, and astrophysical plasmas \cite{ji2011}.
The formation of a thin, kinetic-scale, current sheet is the requirement for fast reconnection \cite{daughton09a,cassak05a}, but not all thin current sheets seemingly unstable to reconnection undergo reconnection.
In a realistic situation, on both sides of the current sheet the pressures are not the same (i.e., asymmetric) and the magnetic fields are not anti-parallel (i.e., the magnetic shear angle $\phi\neq180^\circ$), which induces the diamagnetic drifts of electrons and ions in the outflow direction of reconnection.
%Most current sheets in nature have a certain degree of pressure asymmetry and a guide field (i.e., the shear angle, $\phi$, between the magnetic fields that sandwich the current sheet is not $180^\circ$), 
%for instance, those at Earth's magnetopause \cite{trenchi15a,phan13a}. 
%The diamagnetic drift induced by the cross-current-sheet pressure gradient and the guide field can cause the reconnection x-line to drift in the reconnection outflow direction \cite{swisdak03a}, and it is a consensus that a strong drift may hinder magnetic reconnection. 
It is a consensus that a strong drift may hinder magnetic reconnection \cite{rogers95a, zakharov93a, zakharov92a, biskamp81a,grasso12a, porcelli91a, galeev85a, drake83a, ara78a,beidler11a, swisdak03a}.
However, the fundamentals of how the suppression of reconnection occurs remains unclear, especially in collisionless plasmas.
%The formation of a thin current sheet is the requirement for fast reconnection, where a strong pressure gradient across the sheet may also develop.\\
%Sawtooth crashes in fusion devices are caused by reconnection, 
%Mystery of Sawtooth crash, sudden onset, rapid growth and incomplete reconnection.\\
%Magnetic reconnection serves as an efficient mechanism of transferring magnetic energy into particle energy. 
%It is equally important to know how to shut reconnection off in a current sheet.
%It remains fast in collisionless plasmas, and it remains un-achievable to find a regime with a slower rate. Hence a suppression mechanism may reveal some secret about rate. 
%Diamagnetic drift was proposed to be efficient in suppress reconnection.\\
%Linear vs. nonlinear arguments.

The diamagnetic suppression of tearing modes (i.e., a linear mode that spontaneously leads to reconnection) \cite{rogers95a, zakharov93a, zakharov92a, biskamp81a,grasso12a, porcelli91a, galeev85a, drake83a, ara78a}, among many other ideas, was extensively explored in order to model the sawtooth crashes that spoil the magnetic confinement in fusion devices \cite{von_goeler74a, kadomtsev75a}. In particular, researchers want to explain why the reconnection associated with these crashes usually does not completely consume the available flux \cite{soltwisch92a, levinton94a, yamada94a} after an abrupt onset \cite{zakharov92a,wesson91a}, which suggests a suppression of reconnection. The suppression is reflected in the reduction of the growth rate of tearing modes by the diamagnetic drift frequency, $\omega_j^*$. For instance, a suppression criterion $\omega_j^*> \gamma_0$  is observed through examining the saturation of magnetic islands in two-fluid models \cite{rogers95a,biskamp81a}. Here $\gamma_0$ is the growth rate of tearing modes without the drift effect, and $j=i, e$ for ions and electrons respectively.

On the other hand, recent work examines the nonlinear development of reconnection \cite{beidler11a, swisdak03a} by comparing the diamagnetic drift speed, ${\bf U}^*_j= c{\bf B}\times \nabla {\bf P}_j/(n q_j B^2)$, with the characteristic Alfv\'en speed, $V_A$, which is the typical reconnection outflow speed in non-drifting cases \cite{petschek64a}. Swisdak et al.~\cite{swisdak03a} performed the first particle-in-cell (PIC) simulations to study this drift effect in collissionless reconnection. They observed that the x-line drifts with electrons at a speed $\sim U_e^*$. In this letter, velocity symbols in unbolded font are in the reconnection outflow direction (i.e., $\hat{x}$) unless otherwise specified. Due to the opposite sign of charges, ions drift against electrons and the x-line motion. This fact led them to hypothesize a suppression mechanism that is satisfied when the drift speed of ions in the x-line frame is larger than the Alfv\'en speed, i.e., $|U_e^*-U_i^*| > V_A$. In this scenario, they argued that one of the outflow jets can not develop, hence reconnection is suppressed. Swisdak's criterion is widely used to interpret the occurrence distribution of reconnection at the magnetosphere of Earth \cite{trenchi15a,phan13a} and Saturn \cite{fuselier14a,masters12a}, and in the solar wind \cite{phan10a,gosling13a}.
However, recent gyro-kinetic simulations \cite{kobayashi14a} were conducted to test this criterion in the low shear angle $\phi \ll 1$ (i.e., very strong guide field) and low plasma-$\beta$ ($\sim 0.1-0.01$) limit, and conclude that this suppression condition is not generally satisfied.  In these simulations, the pressure gradient is decomposed into a combination of density and temperature gradients, $\nabla P_j = T_j \nabla n+ n\nabla T_j$. The discrepancy was attributed to the coupling of reconnection with additional instabilities driven unstable by the density \cite{rogers05a}, electron temperature \cite{kobayashi14a, dorland00a} or ion temperature \cite{connor94a} gradients. %such as drift wave instability, mirco-tearing instability and ion-temperature gradient (ITG) modes.

In this letter, we describe in details the development and suppression of reconnection in the regime with a drift velocity $|U_e^*-U_i^*| \gg V_A$, and a moderate magnetic shear angle $\phi\sim 90^\circ$. This is the stable limit of Ref.~\cite{swisdak03a}, and 
%This parameter regime is more relevant to the Earth's magnetopause, as opposed to the $\phi \ll 1$ regime explored using gyro-kinetic simulations \cite{kobayashi14a}. 
this limit requires the difference of plasma-$\beta$ on both sides of the current sheet to be large \cite{swisdak10a}, i.e., $\Delta \beta \gg 1$. In the two cases discussed, reconnection under a drift supported by $\nabla n$ is more or less suppressed, as expected, but reconnection persists under a drift supported by $\nabla T$. %In a fully kinetic simulation, we can evaluate the generalized Ohm's law in a first-principle manner. 
%We identify that both the anisotropy and the nongyrotropy of the full pressure tensor allow the slippage between the electrons and magnetic flux at the outflow region, 
We find that the divergence of the full pressure tensor, $\nabla \cdot {\bf P}_e$, in the generalized Ohm's law allows the slippage between the electrons and magnetic flux at the outflow region. This facilitates reconnection and even results in fast reconnection that lacks one of the outflow jets. 
Through comparing these two cases, we further demonstrate that the asymmetric nature of the current sheet in the $\nabla n$-case slows down the drift of the x-line, leading the following flux to run over and swallow the x-line. A lower bound for this new suppression mechanism is proposed, and it is different from that suggested in Ref.~\cite{swisdak03a} since the drift of ions are not required.

%Through comparing these two cases, we further identify the importance of the local reduction of diamagnetic drift around the x-line, $\Delta U_e^*$, caused by the asymmetric nature of the current sheet. A large $\Delta U_e^*$ leads the following flux to run over and swallow the x-line. A lower bound for the suppression of reconnection, $\Delta U_e^* > V_A$, is proposed. This suppression mechanism does not require the drift of ions and hence is different from that proposed in Ref. \cite{swisdak03a}.

%Here we examine the high $\beta$ limit with $U_j^* \gg V_A$ and guide field $\sim O(5 B_x)$. Ions are not well-magnetized and does not response to the dynamics we describe here. Electrons play the dominant dynamical role.

\begin{table}[ht]
\caption{Asymmetric feature of Runs } % title of Table
\centering % used for centering table
\begin{tabular}{c c c c c c c c c c} % centered columns (4 columns)
\hline\hline %inserts double horizontal lines
Case & $T_1$ & $T_2$ & $n_1$ & $n_2$ & $B_{1y}$ & $B_{2y}$ & $\rightarrow$ $\beta_1$ & $\beta_2$ & $|U_e^*-U_i^*|$  \\ [0.5ex] % inserts table
%heading
\hline % inserts single horizontal line
$\nabla T$ & 12.1 & 1.21 & 1 & 1 & 0.2  & 4.65 & $\rightarrow$ 23 & 0.1 & 7 \\ % inserting body of the table
$\nabla n$ & 6.64 & 6.64 & 1.8 & 0.18 & 0.2 & 4.65 &  $\rightarrow$ 23 & 0.1 & 7 \\
\hline
\end{tabular}
\label{table:nonlin} % is used to refer this table in the text
\end{table}

{\it Simulation setup--}
Kinetic simulations were performed using the particle-in-cell code -{\it VPIC} \citep{bowers09a}. 
The asymmetric configuration employed has the magnetic profile, ${\bf B}=B_{x'}[0.5+\alpha_1\mbox{tanh}(z/\lambda)]\hat{\bf x}'+B_{y'}\hat{\bf y}'$, density $n=\alpha_3[1-\alpha_2\mbox{tanh}(z/\lambda)]$, temperature $T=[\alpha_4-B_x^2/2]/n$ to satisfy the force balance. The pressure gradient is the same for the two cases considered here and the $\nabla T$ ($\nabla n$)-case has an asymmetric temperature (density) profile in the z-direction. We initialize $B_{y'}=0.22 B_{x'}$, $\alpha_1=0.55$, $\alpha_3=0.55$ and $\alpha_4=0.37$. We use $\alpha_2=0.82$ for the $\nabla n$-case and $\alpha_2=0$ for the $\nabla T$-case. To make a better comparison with the runs in Ref.~\cite{swisdak03a}, the 2D simulation plane is then rotated (with respect to the z-axis) to the plane where the reconnecting components are equal, $|B_{1x}|=|B_{2x}|=B_0$. Here the subscripts ``2'' and ``1'' indicate the side at $z > 0$ and $z < 0$ respectively. The resulting asymmetries are summarized in Table I. The magnetic fields are normalized to the reconnecting field $B_0$ and the guide field strength at the location of $B_x=0$ is $\sim 2$. Densities are normalized to $n_0$. Spatial scales are normalized to the ion inertial length $d_i\equiv c/\omega_{pi}$, where the ion plasma frequency $\omega_{pi}\equiv(4\pi n_0 e^2/m_i)^{1/2}$. Time scales are normalized to the ion gyro-frequency $\Omega_{ci}\equiv eB_0/m_i c$. 
%$V_A/c=1/80$ $d_e \equiv c/\omega_{pi}$. . 
%The default unit for velocity is $V_{A0}=B_0/\sqrt{4\pi n m_i}$.
For these two cases, the effective Alfv\'en speeds \cite{swisdak07a} $V^2_{A}=(|B_{1x}|+|B_{2x}|)/[4\pi m_i \left(n_1/|B_{1x}|+n_2/|B_{2x}|)\right]$ are designed to be the same. This $V_A=d_i\Omega_{ci}$ hence will be used to normalize velocities. The temperature is normalized by $m_i V_A^2$. 
%Roughly speaking, $|U^*_i-U^*_e|/V_A \sim (d_i/\lambda)(\Delta \beta)b_g$, where $\Delta \beta \equiv (P_2-P_2)/[0.5(B_{2y}+B_{1y})/8\pi]^2$ and $b_g\equiv 0.5(B_{2y}+B_{1y})/B_x$. For space plasmas, $b_g\sim O(1)$ and $\lambda/d_i \sim O(1)$. Hence $|U^*_i-U^*_e|/V_A \gg 1$ limit requires a $\Delta \beta \gg 1$ plasma.
The initial thickness of the current sheet $\lambda=0.37d_i$, the temperature ratio is $T_i/T_e=2$, the mass ratio is $m_i/m_e=200$, the ratio of electron plasma to gyro-frequency is $\omega_{pe}/\Omega_{ce}=4.17$, and the total magnetic shear angle $\phi=\mbox{tan}^{-1}(B_0/B_{1y})+\mbox{tan}^{-1}(B_0/B_{2y})\approx 90.8^{\circ}$. 
The system size is $76d_i \times 19 d_i$ with grids $8192\times4096$ for the $\nabla T$-case and $4096\times2048$ for the $\nabla n$-case. The resolution of the $\nabla T$-case is higher in order to reduce the noise in the calculation of the generalized Ohm's law presented here. The boundary conditions are periodic in the x direction, while the z direction is conducting for fields and reflecting for particles. We use a localized perturbation to induce a single x-line at $x=0$. Both cases have an initial peak $|U^*_i-U^*_e|/V_A \approx 7$ and $\Delta\beta \equiv |\beta_1-\beta_2|\approx 23$. %and as pointed out in Swisdak et al. \cite{swisdak10a}, with a guide field $\sim O(1)$, the $|U^*_i-U^*_e|/V_A \gg 1$ limit requires a $\Delta \beta \equiv |\beta_1-\beta_2| \gg 1$ plasma.

%Densities are normalized by $n_0$, time is normalized by the ion gyro-freqency $\Omega_{ci}=eB_0/m_i c$, velocities are normalized by the Alfv\'enspeed $V_A\equiv B_0/(4\pi n_0 m_i)^{1/2}$, and spatial scales are normalized by the ion inertia length $d_i\equiv c/\omega_{pi}$. The electron inertia length $d_e \equiv c/\omega_{pi}$ will also be used. $\lambda=0.5d_i$.

\begin{figure}
\includegraphics[width=7cm]{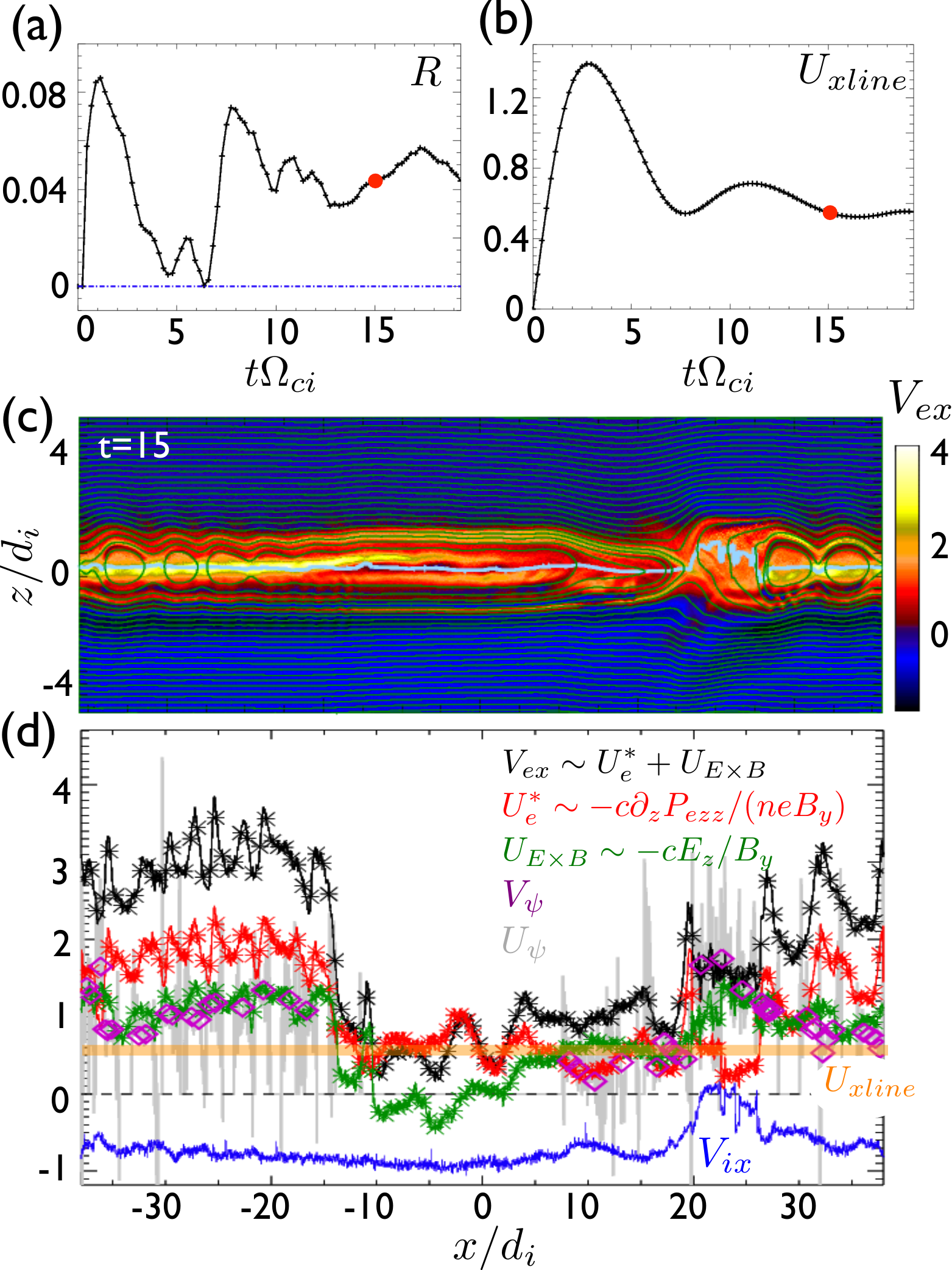} 
\caption {Features of the $\nabla T$-case. In (a), the normalized reconnection rate, $R$. In (b), the drift velocity of the primary x-line, $U_{xline}$. In (c), the x-component of electron velocity with the in-plane magnetic flux overlaid at time $15/\Omega_{ci}$. In (d), the $V_{ex}$, $V_{ix}$, $U_e^*$, $U_{{\bf E}\times{\bf B}}$, $V_\psi$ and $U_\psi$ along the trajectory where $B_x=0$ (indicated as the light blue curve in (c)). The approximation of $V_{ex}$, $U_e^*$ and $U_{{\bf E}\times{\bf B}}$ are plotted as asterisks. The $U_{xline}$ is plotted as a horizontal line for reference.} 
\label{dT}
\end{figure}

{\it $\nabla T$ case--}
%--A contour example of Swisdak.\\
%--Introduce the drifts of plasmas and flux.\\
%--One of the jet does not develop. Non-conventional picture of reconnection. Sharply contrasts with the idea in Swisdak. \\
%--Reconnection rate still $\sim 0.1$! \\
%(Why does $\nabla T$ case have more slippage?? presumably because one side is hotter. ?)
%(Can you have a case that drifts fast and still have reconnection??) 
Contrary to the heuristic prediction proposed in Ref.~\cite{swisdak03a}, reconnection proceeds under such a high $|U_e^*-U_i^*|/V_A$. A reconnection rate $R\sim 0.05$ is sustained, and the x-line drifts in a rather steady velocity $U_{xline}\sim0.5 V_A$ as shown in Fig.~\ref{dT}(a)-(b). Here $R\equiv \left<\partial \Delta \psi/\partial t \right>/(B_0 V_{A})$ with $\Delta \psi \equiv \mbox{max}(\psi)-\mbox{min}(\psi)$ along the $B_x=0$ trajectory and $\psi$ is the in-plane magnetic flux. The $V_{ex}$ at time $15/\Omega_{ci}$ overlaid with the contour of $\psi$ is shown in (c).  
We can understand the composition of $V_{ex}$ from the momentum equation
$m_end{\bf V}_e/dt=-en({\bf E}+{\bf V}_e\times{\bf B}/c)-\nabla\cdot {\bf P}_e$. We consider a subsonic regime where the $d {\bf V}_e/dt$ term is negligible \citep{goldston95a}. By curling the rest of this equation with ${\bf B}$ leads to
$-en[{\bf E}\times{\bf B}-{\bf V}_eB^2/c+{\bf B}({\bf V}_e\cdot {\bf B})/c]-\nabla\cdot {\bf P}_e\times {\bf B}=0$, which shows that the perpendicular flow is the combination of the ${\bf E}\times {\bf B}$ drift and diamagnetic drift: 
${\bf V}_{e\perp}\approx c{\bf E} \times {\bf B}/B^2-c{\bf B}\times \nabla \cdot {\bf P}_e/(enB^2)$. %Here the full electron pressure tensor is considered.

We plot the x-component of electron drifts along the $B_x=0$ trajectory in Fig.~\ref{dT}(d). These velocities along this trajectory are perpendicular to the local magnetic field, hence are relevant to the transport of magnetic structures inside the current sheet. To good approximation, with $B_x=0$ 
\begin{equation}
V_{ex}={V}_{e\perp,x}\approx -\frac{cE_z}{B_y}-\frac{c\partial_z P_{ezz}}{n e B_y}.
\label{Alfven}
\end{equation} 
These approximations are verified by the fact that those asterisks closely trace the solid curves in Fig.~\ref{dT}(d). The $E_z$ associated with the $U_{{\bf E}\times {\bf B}}$ develops in our simulations within time $1/\Omega_{ci}$ partially because the current sheet relaxes into a kinetic equilibrium. 
%There are several possibilities for the slow down of the local drift velocities near the x-line, including the flattening of $\nabla T$ by the heat flux or by the open-up of the x-line geometry, the magnetization of ions, or momentum conservation when the upstream plasmas flow in. 
Note that, although the local drift speed near the x-line drops to a steady value $V_{ex}\sim V_A$ during the nonlinear evolution, reconnection proceeds with a higher $V_{ex} \gtrsim 3 V_A$ in the earlier stage. (e.g., also the tearing modes at $x < -15d_i$ in Fig.~\ref{dT}(c)). 
\begin{figure}
\includegraphics[width=8cm]{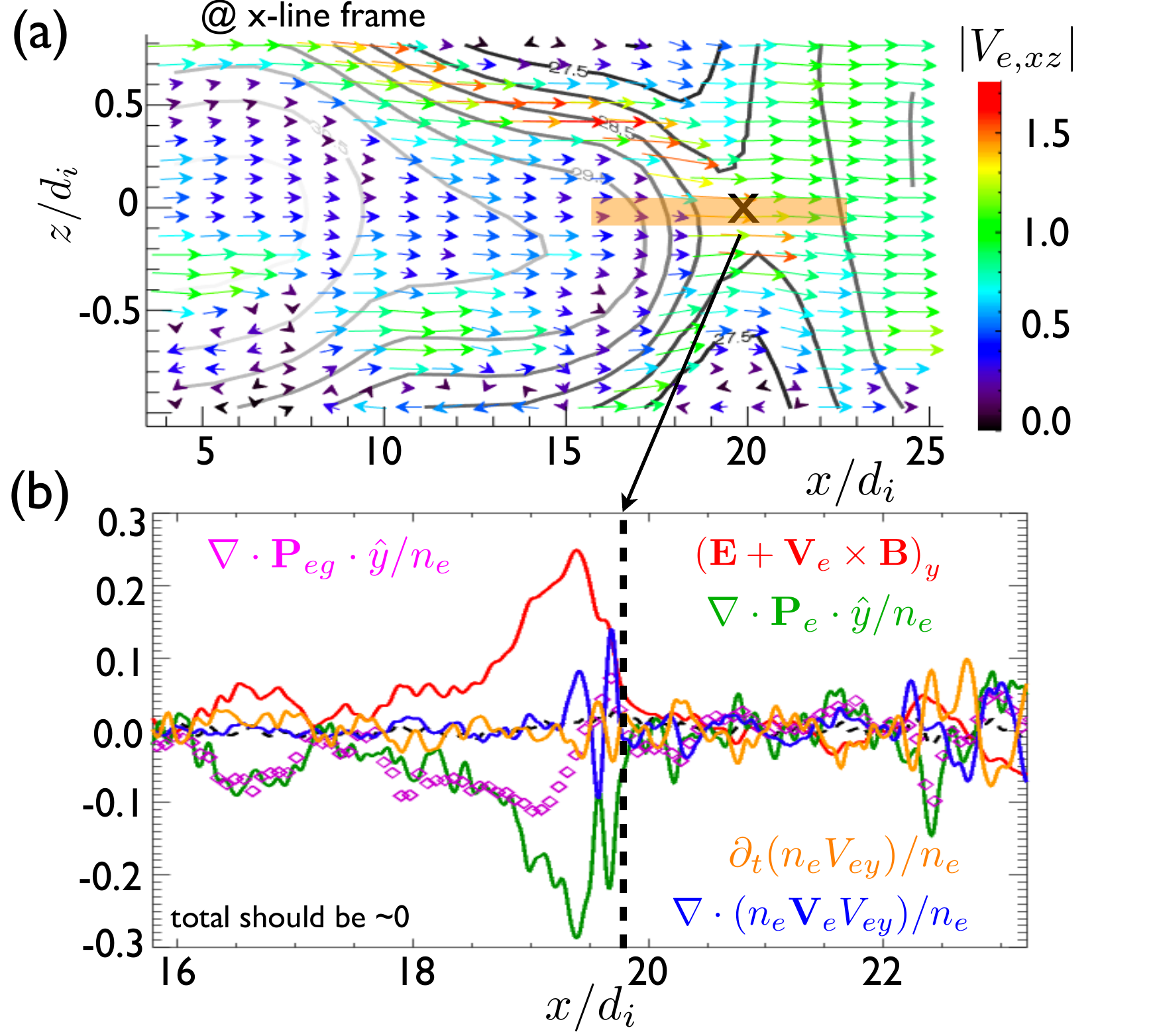} 
\caption {A blow up of the drift x-line in Fig.~\ref{dT}. In (a), the electron flow and the in-plane flux are shown in the x-line drift frame. Here $|V_{e,xz}|\equiv(V_{ex}^2+V_{ez}^2)^{1/2}$. In (b), terms of the generalized Ohm's law are evaluated along the orange line at a fixed z indicated in (a).} 
\label{ohms}
\end{figure}
%This relation is verified in Fig~\ref{}. $U_{{\bf E} \times {\bf B}}=(E_yB_z-E_zB_y)/B^2\sim -E_z/B_y$ since large scale parallel electric field cannot be maintained $E_\| \sim 0$ and $B_y \gg B_z$. A localized $E_z$ establish the electrostatic potential $\phi(x)$ that causes charge separation.\

We zoom in on the region around the x-line in Fig.~\ref{ohms} (a).  The electron flow pattern is shown in the x-line drift frame. Unlike the typical flow pattern around an x-line, the electrons are not repelled leftward of the x-line to form a jet. A similar observation applies for the ions, since $V_{ix} < U_{xline}$ persists inside the current sheet as indicated in Fig.~\ref{dT}(d). Surprisingly, reconnection remains fast with only one jet. 
%These suggest that the reconnection rate here can not be modeled by a argument based on the development of plasma outflow. 
However, the reconnected magnetic flux needs to leave the x-line to both sides, regardless of the magnitude of transport speed. 
%To see how this could possibly work at the left side of the x-line, we examine the out-of-plane component of the non-ideal electric field in the Ohm's law, which tells the frozen-in condition. 
Fig.~\ref{ohms}(b) shows a finite value of the y-component of the non-ideal electric field (red) at the left side of the x-line, which indicates the slippage between electrons and flux. Hence the flux can propagate out of the x-line while the electrons flow in. This is reflected in the fact that the observed flux advection velocity, $V_{\psi}$, is smaller than $U_{xline}$ at the left side of the x-line (i.e., between $x\sim 5$ to $15 d_i$) in Fig.~\ref{dT}(d). The $\nabla\cdot {\bf P}_e$ term (green) contributes to this non-ideal electric field and hence enables this slippage. We approximate the full pressure tensor in a gyrotropic form ${\bf P}_{eg} \equiv P_{e\perp} {\bf I} + (P_{e\|}-P_{e\perp}){\bf b b}$, where ${\bf b}\equiv {\bf B}/|B|$, the parallel pressure $P_{e\|}\equiv{\bf b}\cdot {\bf P}_e \cdot {\bf b}$ and the perpendicular pressure $P_{e\perp}\equiv [\mbox{Tr}({\bf P}_e)-P_{e\|}]/2$. Then $\nabla\cdot {\bf P}_{eg}\cdot \hat{y}=\nabla \cdot (P_{e\|}-P_{e\perp}){\bf b}\mbox{b}_y$. The evaluation of this term (pink) suggests that the pressure anisotropy contributes significantly to this slippage, while the difference between $\nabla\cdot {\bf P}_e\cdot \hat{y}$ and $\nabla\cdot {\bf P}_{eg}\cdot \hat{y}$ closer to the x-line is attributed to the nongyrotropy \cite{aunai13d,hesse11a}.  
%It is noticeable that an even stronger slippage occurs for these secondary tearing modes and x-lines therein at $x < -15d_i$. 
%The temperature anisotropy may be caused by a heat flux induced by the $\nabla_\| T_e\approx \hat{z}\cdot\nabla T_e$, when the reconnected component $B_z$ is generated.

In addition, we can estimate the advection speed of the in-plane flux.
In 2D, we can write ${\bf B}=\hat{y}\times \nabla \psi + B_y \hat{y}$. By performing
$\hat{y}\times [\partial_t {\bf B}+c\nabla\times {\bf E}=0]$ gives
$E_y=(1/c)\partial_t \psi$. This leads the y-component of the electron force balance, $\hat{y}\cdot [{\bf E}+{\bf V}_e\times {\bf B}/c=-(1/en)\nabla \cdot {\bf P}_{e}]$, at where $B_x=0$ into an advection equation, $\partial_t\psi+U_{\psi}\partial_x\psi=0$. The estimated advection velocity of $\psi$ is
\begin{equation}
%\partial_t\tilde{\psi}+V_{ex}\partial_x\tilde{\psi}=0
U_\psi = V_{ex}-\frac{c\nabla\cdot {\bf P}_e\cdot \hat{y}}{enB_z}.
\label{Vpsi}
\end{equation} 
%(i.e., $\psi$ is constant along the characteristics $x(t)$ that satisfies $dx/dt=U_{\psi}$ and $d\psi/dt=0$). 
This expression is valid at the region where $B_z\neq 0$. This also excludes the singular x-line where $\psi$ is generated and the $\nabla\cdot {\bf P}_e$ term should be treated as a source term.
This $U_\psi$ in Fig.~\ref{dT}(d), although noisy, follows well with the measured $V_\psi$. %If the frozen-in condition is maintained with $\nabla\cdot {\bf P}_e\cdot \hat{y}/n$ being negligible, 
%a perturbed flux propagates at speed $V_{\psi}=V_{ex}\approx U_{{\bf E} \times {\bf B}}+U_e^*$ locally. 
%In this limit, it appears to simply comply with the frozen-in flux condition. However, the velocity of flux could be arbitrary if we try to understand from the general condition of flux preservation \cite{dorelli08a}.  
In this case $U_\psi$ and $V_\psi$ do not match well with $V_{ex}$, which again implies a significant slippage that facilitates reconnection.
%, but less slippage does not directly lead to the suppression. %and hence there is slippage between electrons and flux. %This velocity is confirmed by comparing the $V_{e\perp}$ near the x-line with $V_{xline}$ marked in Fig.~\ref{dT}(d). %It also interesting to note, since the electron inflow does not drift, due to the momentum conservation, the x-line drift velocity tends to drop to order of local $V_{Ae}$. Another possibility of this slow down is due to the opening of reconnection exhausts, which could greatly reduce the pressure gradient.

\begin{figure}
\includegraphics[width=8 cm]{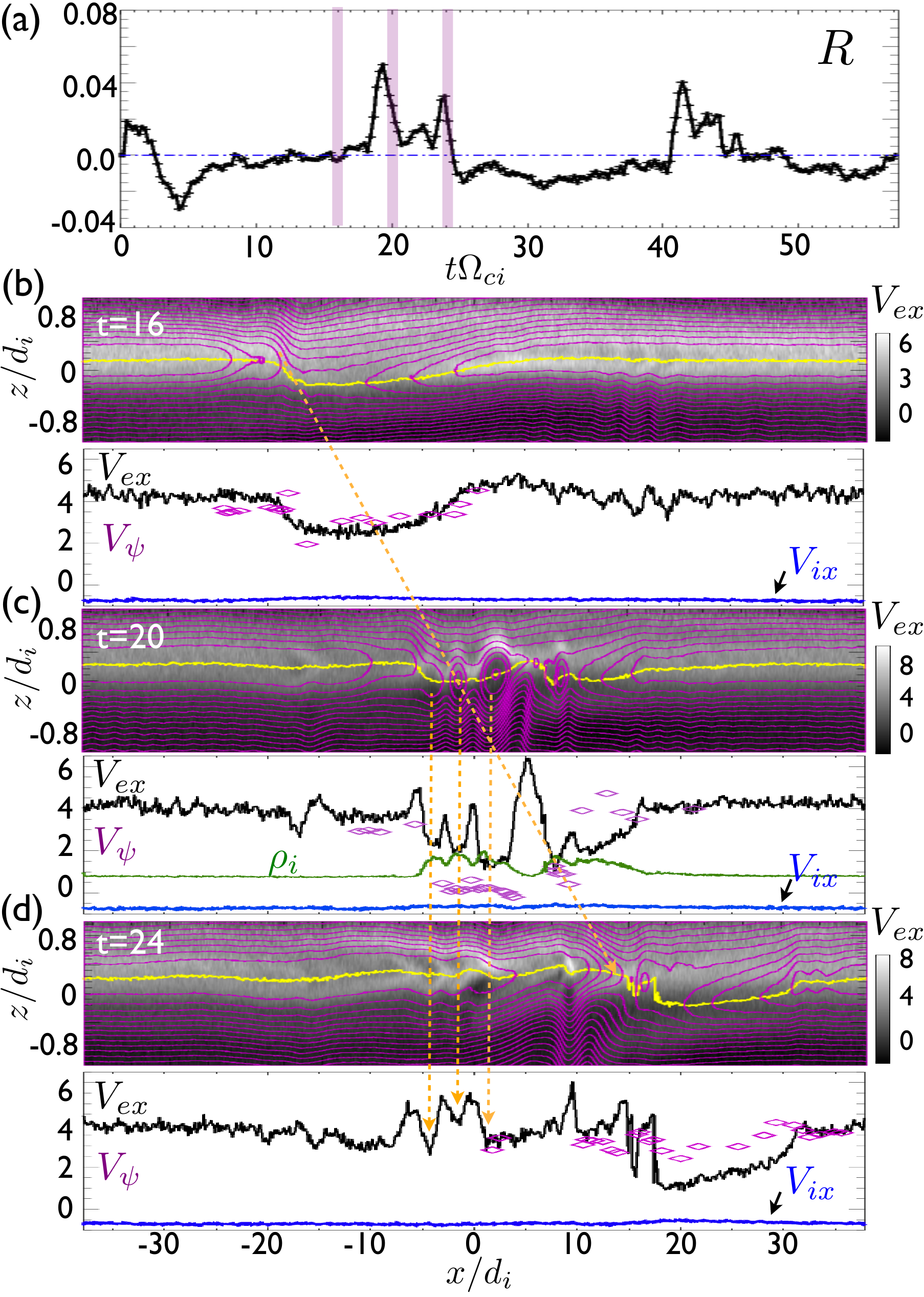} 
\caption {The long term evolution of the $\nabla n$-case. In (a), the normalized reconnection rate, $R$. Panels (b), (c) and (d) show the $V_{ex}$ map overlaid with the in-plane flux at times marked in (a). The $V_{ex}$, $V_{ix}$ and $V_{\psi}$ along the trajectory where $B_x=0$ (the yellow curves in the 2D map) are plotted underneath each panel. The ion gyro-radius, $\rho_i$, is also shown in (c). The oblique yellow dashed line tracks the motion of the following flux, while the three nearly vertical yellow dashed lines track the motion of tearing modes and active x-lines therein.} 
\label{dN}
\end{figure}

{\it $\nabla n$ case--}
%--Illustrate the suppression mechanism.\\
%In particular, the current sheet without the initial perturbation remains stable to tearing instability and reconnection. 
On the other hand, the suppression becomes pronounced in the $\nabla n$-case. 
For the simulations presented here, the induced x-line does not reconnect. However, over longer times tearing modes are triggered and then suppressed, resulting in a periodic burst of reconnection as shown in Fig.~\ref{dN}(a). This cyclic behavior reveals the nature of suppression.

%If like swisdak, the suppression should take a way of head-on collision with the right hand flux.%
Fig.~\ref{dN}(b)-(d) show the contours of $\psi$ and $V_{ex}$ around the period of the first burst of reconnection at time 16, 20 and $24/\Omega_{ci}$. The induced x-line remains inactive in (b). The tearing modes with a significant growth are triggered at the trailing edge of the primary island in (c), which boost the reconnection rate up to $\sim 0.05$. In (d), these tearing islands do not propagate and hence are run over and swallowed by the following flux that maintains a steady positive velocity $\sim 4 V_A$. This motion completely shuts off the reconnection. The tendency is also reflected in the the profile of $V_\psi$ (purple), which follows the trend of $V_{ex}$. 
%Since the measurement of $V_{\psi}$ is possible only at regions where the variation of $\psi$ is larger than the noise level, the $V_{ex}$ curve is plotted for the guidance. 
Due to the asymmetric nature of the current sheet, a trough in the $V_{\psi}$ profile develops with the x-line sitting at the bottom as shown in Fig.~\ref{dN}(b). The negative slope at the left side of the x-line indicates this tendency of running-over. The birth of tearing modes further diverts the electron flow to the upper side of tearing islands, and this only deepens the $V_{\psi}$ trough in (c), which eventually results in the running-over in (d). The same reason explains why tearing modes only become feasible temporarily at the edge where the $V_{\psi}$ profile has a positive slope, which delays the running-over.
%It is therefore clear that the velocity difference between the immediate left-coming flux and tearing mode (i.e., a negative slope in $V_{\psi}$ profile at left of the x-line) results in the suppression. The same reason explains why tearing modes in the $\nabla n$-case become feasible temporarily at the trailing edge of the primary island where the velocity profile $V_{\psi}\sim V_{ex}$ has a positive slope (Fig.~\ref{dN}(b)). 
%Note, the slow down of tearing mode cannot be explained by the magnetization of ions. Because the $\rho_i\sim ?? d_i$ is similar for both cases but only one case shows the slow down.
%It is also interesting to point out that the burst of reconnection with a similar rate can last longer if tearing modes grow faster and form a spatially longer tearing-mode-chain, so that the left-coming flux needs more time to swallow them one-by-one. 
In contrast, for the $\nabla T$-case shown in Fig.~\ref{dT}(c), the current sheet and the resulting tearing modes are more or less symmetric. Hence the tearing islands and the induced x-line can pick up the peak drift velocity available to advect flux. The in-plane magnetic structures in the $\nabla T$-case all propagate in a positive velocity $\sim V_A$ as indicated by the $V_{\psi}$ in Fig.~\ref{dT}(d). 
These x-lines will not be run over by the following flux and reconnection persists. 
%As opposed to the asymmetric feature of tearing modes, a potential magnetization of the ions does not seem to cause the different outcome: For both cases the tearing modes grow to a similar amplitude and both the gyro-radius and drift speeds of ions are similar, but only the $\nabla n$-case significantly slow down the drift of tearing modes. 
As opposed to the asymmetric feature of tearing modes, a potential magnetization of the ions does not cause the local reduction of $V_\psi$ in the $\nabla n$-case, since the local ion gyro-radius ($\sim d_i$) does not decrease at the tearing islands and the ion drift $V_{ix}$ is barely affected as shown in Fig.~\ref{dN}(c). These observations suggest that this suppression mechanism does not require the drift of ions, and hence differs from that proposed in Ref.~\cite{swisdak03a}.
% Fact1: Since the linear tearing stage where ions do not affect yet, $\nabla T$-case drifts, $\nabla n$-case does not drift.
% Fact2: A linear tearing in the $\nabla n$ case is totally suppressed before ions have the chance to get involved.
% Fact3: If like Swisdak, then ions should cause a head-on collision.
% Fact4: You could use $m_i/m_e=1836$ EMHD limit to test this.?
% Fact5: The flux does not propagate with ions, hence ions are likely not frozen-in for the whole regime.
% Fact6: Electron should also get more magnetized locally inside the island.
% Fact7: In principle, Ion diamagnetic drift should not cause different drift speed at two side of x-line, since it is not a guiding-center drift!
It is also interesting to point out that the burst of fast reconnection can last longer if tearing modes grow faster and form a spatially longer tearing-mode-chain, so that the following flux needs more time to swallow them one-by-one. This phenomenon is seen in a companion case with hotter electrons (e.g., a case with $T_i/T_e=0.3$, not shown).

%In particular, the flatness of the $V_{\psi}$ profile in the region with tearing modes at $x<-15d_i$ in Fig.~\ref{dT}(d) does not look like those local troughs caused by tearing modes in Fig.~\ref{dN}(d). 
%For instance, in a similar case with smaller $T_i/T_e$  (not shown) where electrons are hotter to make tearing growth more rapidly. Note, the degree of slippage does not determine suppression directly, it just makes $V_\psi$ depart from $V_{ex}$. It is the $\Delta V_\psi$, that leads to suppression. The slippage does, however, helps the reconnection in the $\nabla T$ case \\

%\begin{figure}
%\includegraphics[width=8cm]{dT_evolve} 
%\caption {Go back to the original question: Why $\nabla T$ is unstable? Symmetric tearing in $\nabla T$-case are convected along with the ambient jets, hence not suppressed.} 
%\label{dT_evolve}
%\end{figure}

\begin{figure}
\includegraphics[width=8.5cm]{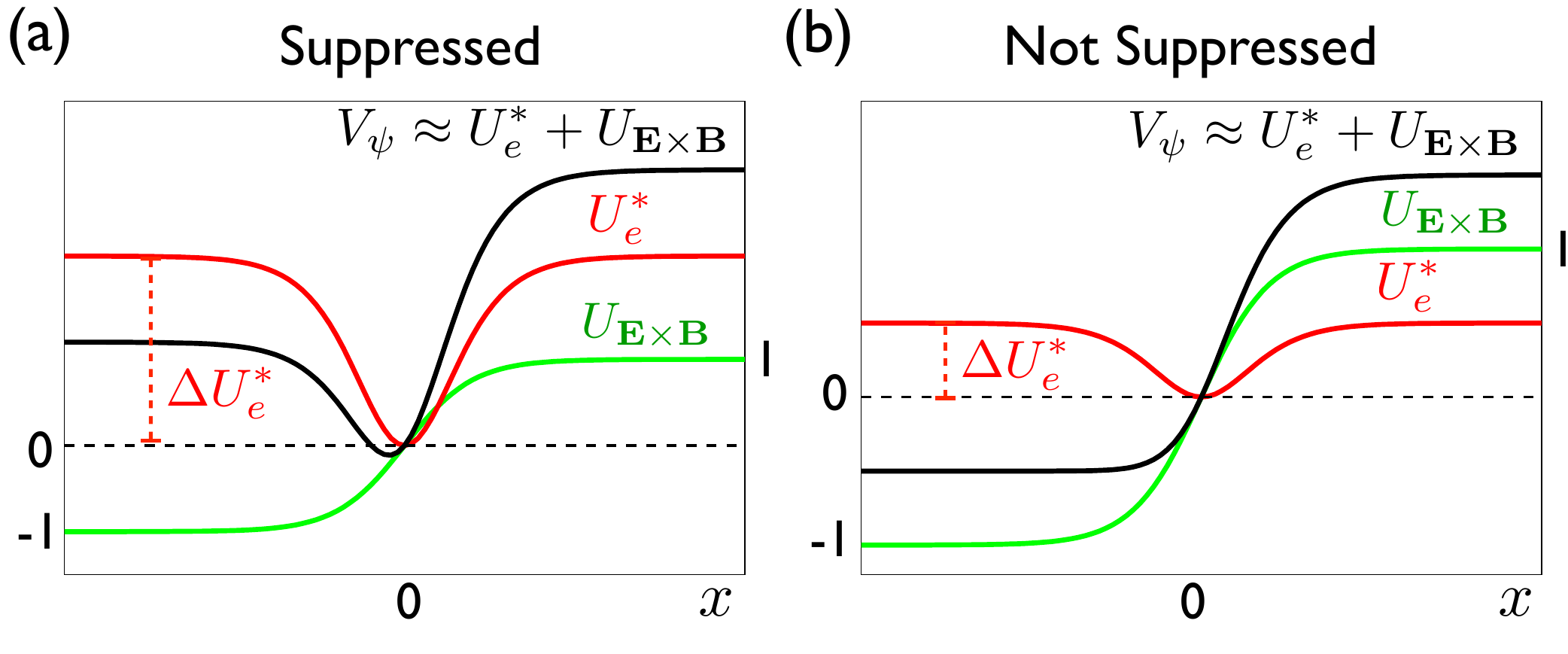} 
\caption {The velocity profiles around the x-line in the x-line drift frame. Here the x-line is at $x=0$ and the slippage is assumed negligible. In (a), $\Delta U^*_e > V_A$. In (b), $\Delta U^*_e < V_A$.} 
\label{sup_model}
\end{figure}

{\it Model--}
To model the lower bound for this suppression, we consider an ideal situation without slippage, i.e., $V_\psi \approx V_{ex}\approx U_e^*+U_{{\bf E}\times{\bf B}}$. The asymmetric nature of the current sheets causes a trough in the profile of $U_e^*$ locally near the x-line, while the magnetic tension force in $d_i$-scale tends to drive electrons to the Alfv\'en speed $V_{A}$ outwardly from the x-line \cite{rogers01a}, and this effect enters the $U_{{\bf E} \times {\bf B}}$ drift. %On the other hand, the suppression is not expected if the magnetic tension associated with the reconnected field is strong enough \cite{swisdak03a}.    %It is not the ion-Alfv\'en speed in such a high $\beta$ plasmas because ions are not well-magnetized in the electron-scale. 
%(i.e., To derive the Alfven speed in a well frozen-in plasma, we $\nabla \times$ the momentum equation to get rid off the $\nabla P$ term. Hence the plasma response to the tension force is totally reflected in $U_{E\times B}$).  
%In Fig.~\ref{sup_model}, we summarize our understanding.
Fig.~\ref{sup_model}(a) shows a case where the depth of the trough, $\Delta U^*_e$, is larger than the asymptotic value of $|U_{{\bf E} \times {\bf B}}|\sim V_A$. Note that, these velocities are measured in the x-line drift frame. The resulting $V_{\psi}$ has a trough with a negative slope at the left side of the x-line, hence the x-line will be run over. If $\Delta U^*_e < V_{A}$ as in (b), the $V_{\psi}$ profile only has a positive slope and is free from this suppression mechanism.
%favoring the development of tearing modes, and is free from this suppression mechanism. 
The $\nabla n$-case falls into the category of Fig.~\ref{sup_model}(a). The $\nabla T$-case falls into the category of (b) after accounting for the slippage.  
%, even though the $V_{ex}$ profile near the x-line in Fig.~\ref{dT}(d) weakly resembles Fig.~\ref{sup_model}(a) because of the slippage. 
Based on these observations, we conclude that the suppression of reconnection, a least, requires $\Delta U^*_e > V_{A}$.

{\it Summary and Discussion--}
A new suppression mechanism cased by the local reduction of flux transport speed is identified inside the ion diffusion region. This robust mechanism can completely shut off reconnection after a rapid onset. 
The magnitude of $\Delta U^*_e$ is determined by the asymmetric nature of the current sheet and the resulting tearing modes. For the $\nabla n$-case, the asymmetric nature may come from the mismatch between the x-line and the flow stagnation point (i.e., $\delta_ {x2}/\delta_{x1}= B_2/B_1$ and $\delta_{s2}/\delta_{s1}= n_2 B_1/n_1 B_2$ in Ref.~\cite{cassak07b}), which does not develop with only the asymmetry in temperature, as the $\nabla T$-case. 
%It requires more work to carefully model the $\Delta U_e^*$ in a general condition.
%A rough estimation will be $\Delta U^*_e \sim 2|\rho_1 B_2^2-\rho_2 B_1^2|/[(B_1+B_2)(\rho_1B_2+\rho_2B_1)] U^*_{e,peak}$. 
In the limit of strong density asymmetry, the tearing modes and active x-lines therein barely drift as in the $\nabla n$-case presented here, then the $\Delta U^*_e$ is similar to the equilibrium $U_e^*$ of the current sheet. The lower bound for this suppression reduces to a simpler form $U_e^*> V_A$.

%The higher degree of slippage in the $\nabla T$-case may partially come from the heat flux induced by the field-aligned $T_e$-gradient, which arises when $B_z$ is generated. The slippage enables reconnection with only one jet in the x-line drift frame. 
This study also shows that reconnection with only one jet in the x-line drift frame is possible due to the slippage between plasmas and the magnetic flux.
This fact suggests the need for improving the identification of reconnection events at Earth's magnetopause, and this could be studied using NASA's on-going Magnetospheric Multiscale mission that is capable of evaluating $\nabla\cdot {\bf P}_e$ routinely.

\acknowledgments
Y. -H. Liu thanks for helpful discussions with J. F. Drake, M. Swisdak, P. Cassak, W. Daughton and A. Spiro. This research was supported by an appointment to the NASA Postdoctoral Program at the NASA-GSFC, administered by Universities Space Research Association through a contract with NASA. Simulations were performed with NASA Advanced Supercomputing and NERSC Advanced Supercomputing.  

%\bibliography{paper}

\newpage

\end{document}